\documentclass[preprint,onecolumn,nofootinbib]{revtex4}
\usepackage{graphicx}
\pdfoutput=1
\usepackage[colorlinks=true,linkcolor=blue,urlcolor=blue,filecolor=black,citecolor=red,pdfstartview=FitV,pdftitle={},pdfsubject={},pdfkeywords={},pdfpagemode=None,bookmarksopen=true]{hyperref}
\usepackage{epsfig}
\usepackage{amsmath}
\usepackage{amsfonts}
\usepackage{amssymb}
\usepackage{color}%
\usepackage{dcolumn}
\providecommand{\U}[1]{\protect\rule{.1in}{.1in}}

\newcommand{\f}{\begin{equation}}
\newcommand{\ff}{\end{equation}}
\newcommand{\fa}{\begin{eqnarray}}
\newcommand{\ffa}{\end{eqnarray}}

\begin{document}
\title{
Holographic Entanglement Entropy Close to Quantum Phase
Transitions}
\author{Yi Ling $^{1,2}$}
\email{lingy@ihep.ac.cn}
\author{Peng Liu $^{1}$}
\email{liup51@ihep.ac.cn}
\author{Chao Niu $^{1}$}
\email{niuc@ihep.ac.cn}
\author{Jian-Pin Wu $^{3,2}$}
\email{jianpinwu@gmail.com}
\author{Zhuo-Yu Xian $^{1}$}
\email{xianzy@ihep.ac.cn}
\affiliation{$^1$ Institute of High
Energy Physics, Chinese
Academy of Sciences, Beijing 100049, China\\
$^2$ State Key Laboratory of Theoretical Physics, Institute of
Theoretical Physics, Chinese Academy of Sciences, Beijing 100190,
China\\
$^3$Institute of Gravitation and Cosmology, Department of Physics, School of Mathematics and Physics,
Bohai University, Jinzhou 121013, China
}
\begin{abstract}
We investigate the holographic entanglement entropy (HEE) of a
strip geometry in four dimensional Q-lattice backgrounds, which
exhibit metal-insulator transitions in the dual field theory.
Remarkably, we find that the HEE always displays a peak in the
vicinity of the quantum critical points. Our model provides the
first direct evidence that the HEE can be used to characterize the
quantum phase transition (QPT). We also conjecture that the
maximization behavior of HEE at quantum critical points would be
universal in general holographic models.

\end{abstract}
\maketitle

\section{Introduction}
Quantum phase transitions (QPT) are believed to give rise to some of the most
interesting phenomena in condensed matter physics~\cite{Sachdev:2000qpt}.
Yet a systematic characterization of QPTs remains an outstanding open question.
 Entanglement could provide an important handle for
understanding QPT. For example, there have been indications that entanglement is enhanced at
a quantum critical point (QCP)~\cite{Osborne:2002zz,Osterloh:2002na,Vidal:2002rm,
YChen:2006jop}(for a review, see~\cite{Amico:2007ag}). Furthermore, change of entanglement patterns
may underlie certain QPTs which do not involve symmetry breaking or traditional order parameters.

QPTs are often very difficult to analyze as they naturally occur in strongly
correlated many-body systems. Entanglement entropy is also notoriously hard to calculate.
The AdS/CFT correspondence provides powerful tools for both analyzing QPTs and
computing entanglement entropy~\cite{Ryu:2006bv,Takayanagi:2012kg} in strongly coupled systems.

In this paper we show that in a class of strongly coupled
holographic systems undergoing QPTs, entanglement entropy attains
a maximum at the corresponding QCPs. Thus entanglement entropy can
be used as a diagnostic for QPT. Behavior of entanglement entropy
near various holographic thermal phase transitions were
investigated before
in~\cite{Albash:2012pd,Cai:2012sk,Arias:2012py,Kuang:2014kha}.

\section{The holographic setup and phase diagram}
Metal-insulator transitions are observed in many condensed matter systems and
often involve strongly coupled physics.
Recently there has been
progress in simulating metal-insulator transitions in holographic
systems in (2+1)-dimension~\cite{Donos:2012js,Donos:2013eha,Donos:2014uba,Ling:2014saa}.
Recall that the radial direction of an asymptotically Anti
de-Sitter (AdS) spacetime can be understood as a geometrization of
the renormalization group (RG) flow for the dual field theory. The IR physics is then encoded in the
geometry of the deep interior of the spacetime, to which we will refer to as the IR geometry.
Thus a metal-insulator transition is realized on the gravity side as a geometric transition: as external parameter(s) are dialed, the IR geometry deforms from that describing a metallic phase to
that corresponding to an insulator. At a finite chemical potential, the IR geometry describing a metallic phase
is well known, given by $AdS_2\times R^2$~\cite{Hartnoll:2012rj}. The key to recent development came from construction of
the so-called Q-lattice models~\cite{Donos:2013eha} which give rise to new IR geometries dual to insulators.

Let us now describe the gravity set up for Q-lattice models.
The Lagrangian of the gravity dual
can be written as
\begin{eqnarray}
\mathcal{L}=R+6-\frac{1}{2}F^{\mu\nu}F_{\mu\nu}-|\nabla
\Phi|^2-m^2|\Phi|^2, \label{eq:action}
\end{eqnarray}
where the $AdS$ length scale is set to unity. $\Phi$ is a complex
field which will be used to introduce a lattice structure along
one of the spatial directions, say $x$, by the ansatz
$\Phi=e^{i\tilde{k}x}\varphi$ with $\varphi$ $x$-independent. A
remarkable feature of such an ansatz is that while translational
symmetry is broken, the gravity equations of motion are still ODEs
instead of PDEs, bringing great simplifications. This construction
is analogous to the construction of Q-balls which is employed to
build spherically symmetric solitons in \cite{Coleman:1985ki},
thus is called the holographic Q-lattices.

The full solution to equations of motion of~\eqref{eq:action} can be written in a form
\begin{equation}
\begin{aligned}
\label{sol}
 & ds^2={1\over
z^2}\left(-P(z) dt^2+\frac{dz^2}{P(z)}+V_1dx^2+V_2dy^2\right),\\
 & A=\mu(1-z)a dt,\qquad
\Phi = e^{i \tilde{k} x}z^{3-\Delta}\phi,
\end{aligned}
\end{equation}
where $P(z)\equiv U(1-z)(1+z+z^2-\mu^2z^3/2)$ and
$\Delta=3/2\pm(9/4+m^2)^{1/2}$. Notice that $U,V_1,V_2,a$ and
$\phi$ are functions of the radial coordinate $z$ only and $\mu$
is the chemical potential of the dual field theory. We set the
boundary condition for $\phi$ as $\phi(0)\equiv\tilde\lambda$,
which is understood as the lattice amplitude.
Then the solutions of the background are specified by three
dimensionless parameters, namely the temperature $\tilde T/\mu$,
lattice amplitude $\tilde\lambda/\mu^{3-\Delta}$, and lattice wave
number $\tilde k/\mu$. For simplicity, we will denote these
quantities in short by $T,\lambda,k$ through this paper. {The
metric has an event horizon at $z=1$ and the spacetime boundary is
at $z=0$.} The Hawking temperature is $T=(6-\mu^2)U(1)/(8\pi\mu)$.
We will work at temperature $T = 0.001$, which is
low enough to observe quantum critical phenomenon.
In addition, we will set the mass of the scalar field as $m^2=-2$,
with a brief discussion on other values of the mass in the end of this paper.

\begin{figure}
\begin{center}
  \includegraphics[scale=0.6]{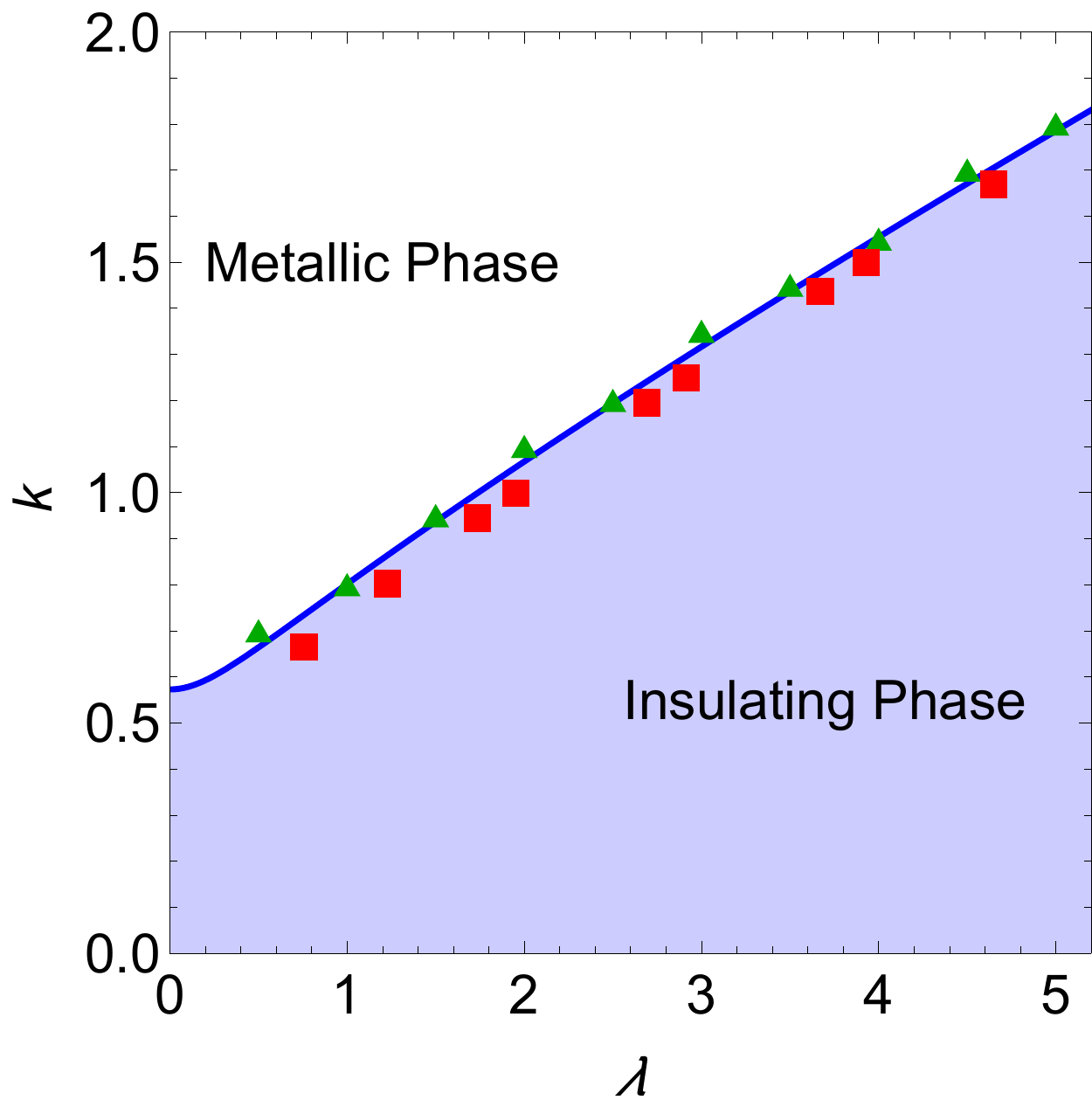}
\end{center}
\caption{The phase diagram over $(\lambda,k)$ plane ($\lambda\neq
0,k\neq 0$) for the metal-insulator transition in Q-lattice
background. The upper and the lower region of the plot corresponds
to metallic and insulating phase, respectively. The HEE with large
$l$ is numerically computed through Eq.(\ref{slform2}). Green
triangles represent the peaks of HEE when varying $k$ but fixing
$\lambda$, while red squares are peaks of HEE when varying
$\lambda$ but fixing $k$.}\label{phase}
\end{figure}

The phase diagram for the metal-insulator transition over Q-lattice
backgrounds~\eqref{sol} can be obtained by examining the DC conductivity
$\sigma_{\text{DC}}$ of the dual field theory along $x$ direction, and
is presented in Fig.~\ref{phase}. $\sigma_{\text{DC}}$ can be computed using standard methods
and it can be written in terms of various geometric quantities in~\eqref{sol} evaluated at the
horizon, i.e.
\begin{equation}
{\sigma_{\text{DC}}=\left.
\left( {\sqrt {\frac{{{V_2}}}{{{V_1}}}}  + \frac{{{\mu
^2}{a^2}\sqrt {{V_1}{V_2}} }}{{{k^2}{\phi ^2}}}} \right)\right|_{z
= 1}.}
\end{equation}
As a practical diagnostic, we classify the dual system into a metallic phase or an insulating
phase by the temperature dependence of DC conductivity around
$T=0.001$. More explicitly, for the metallic phase we expect
$\partial_T\sigma_{\text{DC}}<0$, while for the insulating phase
$\partial_T\sigma_{\text{DC}}>0$. As a result, the QCPs are
characterized by $\partial_T\sigma_{\text{DC}}=0$. They form a critical
line in $k-\lambda$ plane as shown in Fig.\ref{phase}. We stress that this phase
structure has little change when decreasing the temperature further.

\section{HEE and numerical results}
In the  AdS/CFT correspondence
the entanglement entropy for a region $A$ is obtained from gravity
as the area of the minimal surface $\gamma_A$ in the bulk geometry
which ends at $\partial A$~\cite{Ryu:2006bv}, i.e.
\begin{equation}\label{hee}
  S_A=\frac{\text{Area}(\gamma_A)}{4G_{N}},
\end{equation}
where $G_{N}$ is the bulk Newton constant.

We now consider the holographic entanglement entropy (HEE) for a
strip region in the boundary system described by~\eqref{sol}.
We take strip to stretch in $y$-direction with length
$L_y$($\to \infty$) and $x$-direction with width $2l \ll L_y$.
While the translational symmetry along $x$ direction is
broken by $\Phi$, the metric components in~\eqref{sol} are
still functions of $z$. It is then straightforward to find the
minimal surface for the strip region, which can be specified by
the location $z_*$ of the bottom of the minimal surface
 in $z$-direction. We find that $z_*$ satisfies the equation
\begin{equation}\label{slform1}
   l = \mu \int_0^{z_*} dz z^2 \sqrt{ \frac{V_1(z_*) V_2(z_*)} {P(z) V_1(z) W(z_*,z)}},
\end{equation}
where $W(z_*,z)\equiv z_*^4 V_1(z) V_2(z) - z^4 V_1(z_*) V_2(z_*)$.

\begin{figure}
\center{
\includegraphics[scale=0.5]{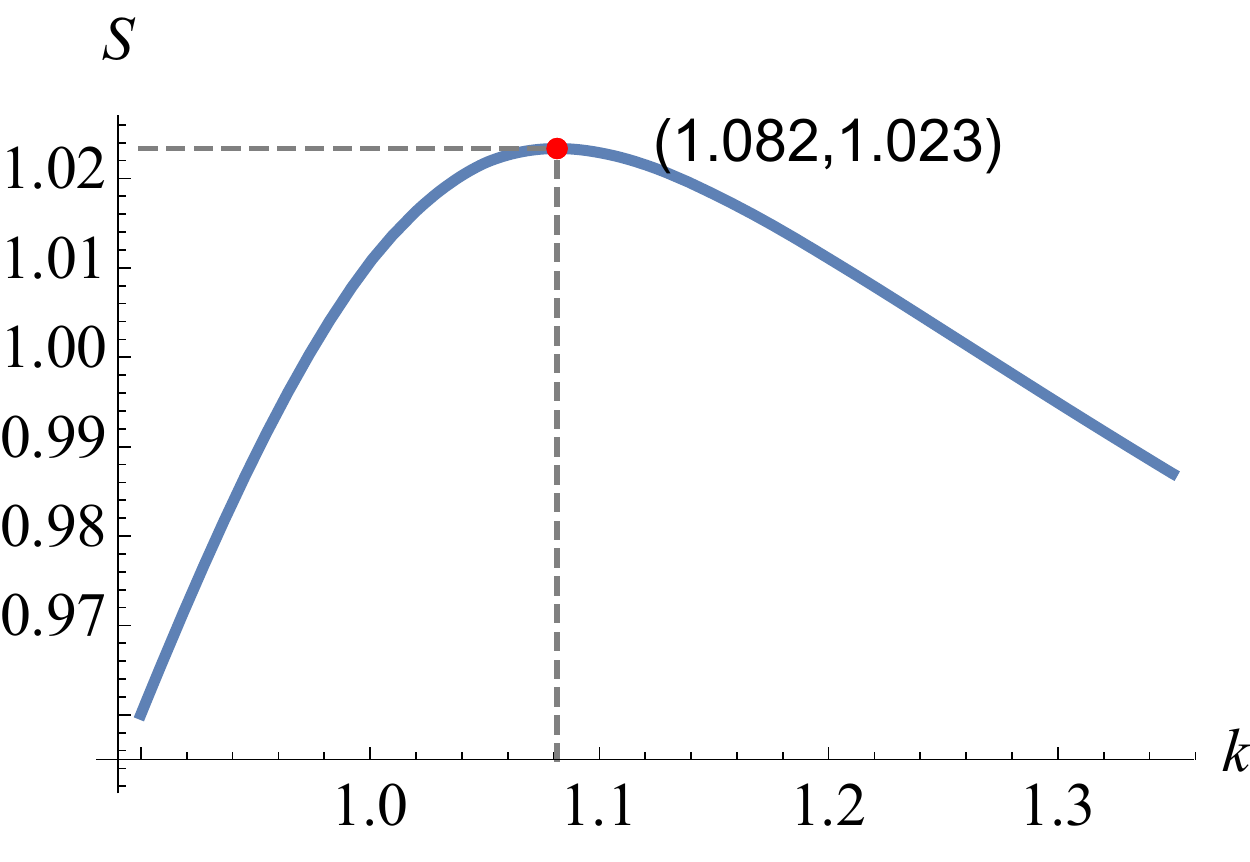}\;\;\;\;
\includegraphics[scale=0.5]{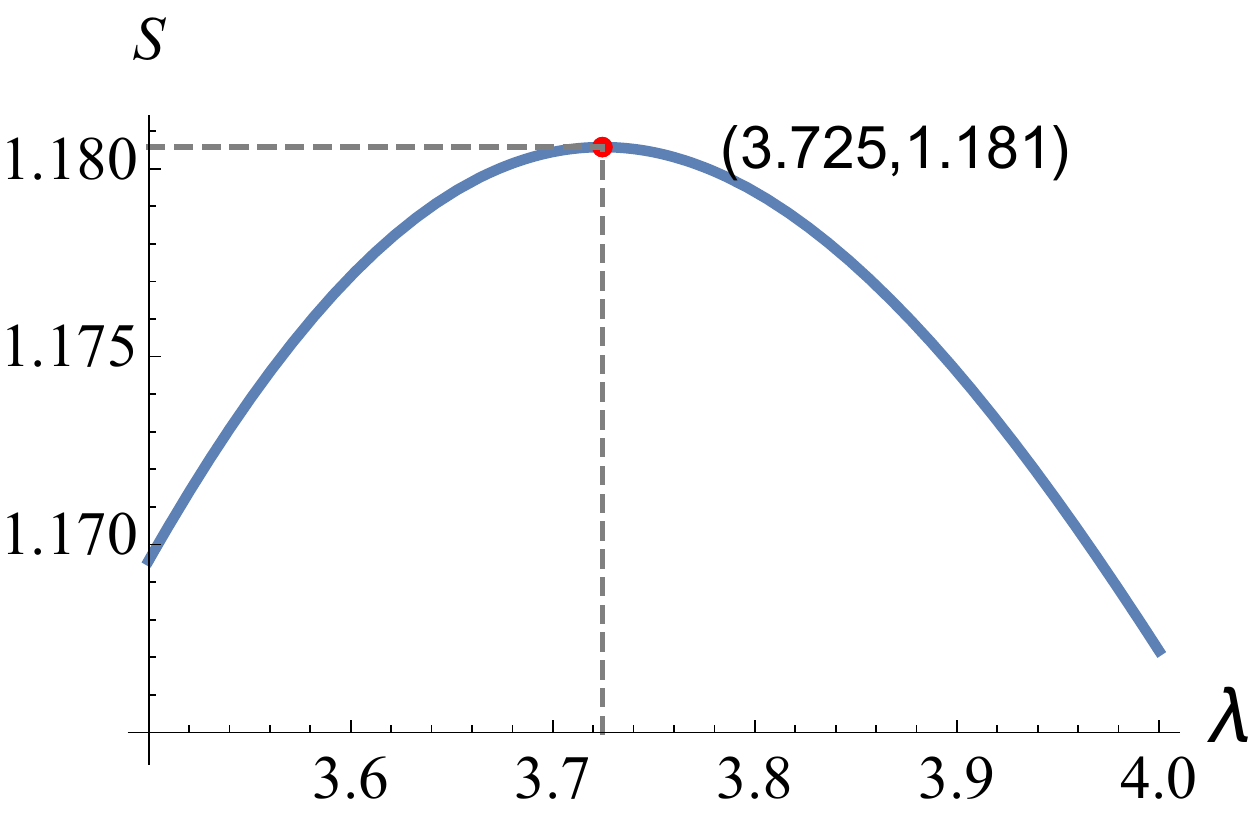}
\caption{\label{klplot} The left plot is for HEE $S$ as a function
of the lattice wavenumber $k$ with $l=5.56, \lambda=2$, while the
right plot is for $S$ as a function of the lattice amplitude
$\lambda$ with $l=5.44, k=1.44$.
 }}
\end{figure}

\begin{figure}
\center{
\includegraphics[scale=0.6]{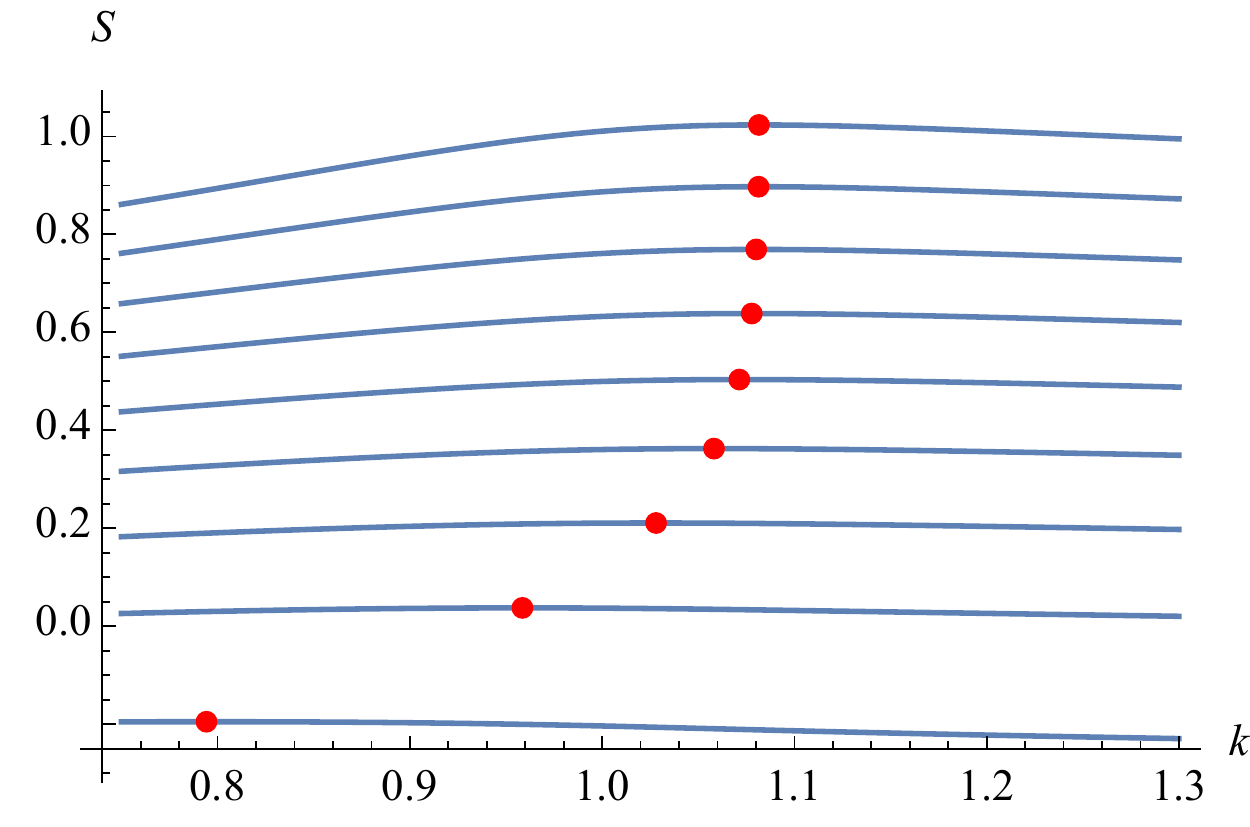}
\caption{\label{trend} The parameter $\lambda$ is fixed at
$\lambda=2$ for the above plot. The turning point approaches the
critical point when the width of the strip is increased from
$0.91$ to $5.56$ uniformly. The maximum of each curve is marked as
a red dot in figure.}}
\end{figure}

From Eq.(\ref{hee}) we find that after subtracting the vacuum part the entanglement
entropy $S$ behaves as
\begin{equation}\label{slform2}
  S = \frac {L_y}{ 2 \mu G_N} \left\{  - \frac1{z_*} + \int_0^{z_*} \frac{dz}{z^2}
   \left[ \frac{z_*^2 V_1(z) V_2(z)}{ \sqrt{P(z) V_1(z) W(z_*,z)}} -1 \right]
   \right\} \ .
\end{equation}
Firstly, we compute the HEE $S$ as a function of the parameter $k$
and $\lambda$ separately when the strip width $2l$ is finite but
fixed at a relatively large value. Our results are illustrated in
Fig. \ref{klplot}. Interestingly enough, we observe that the HEE
displays a pronounced peak in both plots. Moreover, we find that
the location of such turning points is independent of the width of
the strip when $l$ is relatively large. To show this we plot the
shift of the peaks of HEE with the width of the strip in Fig.
\ref{trend}. Evidently, the peaks of the HEE converge to a fixed
point with a definite value of $k$ when the width of the strip is
becoming large enough. The same convergent behavior is also
observed in the case when varying the parameter $\lambda$ with $k$
fixed. Secondly, we mark the turning points of the HEE with large
$l$ in the $(\lambda,k)$ plane of the phase diagram as illustrated
in Fig.\ref{phase}. Remarkably, we observe that all the turning
points of the HEE are distributed in the vicinity of the
trajectory of the critical points, clearly indicating that HEE can
be used to characterize the occurrence of QPTs.

Next, we are interested in the large distance behavior, i.e.
$l \to \infty$. In this limit it can be readily seen
from~\eqref{slform1} that $z_* \to 1$, i.e. the bottom of the
minimal surface approaches the horizon. It then follows
from~\eqref{slform2} that $S$ is dominated by the contribution of
the part of the minimal surface near the horizon. More explicitly,
as $z_* \to 1$ from~\eqref{slform1} and~\eqref{slform2}, both $S$
and $l$ are logarithmic divergent,
\begin{equation}\label{slform}
 \begin{aligned}
S &\approx \left.\frac{V_2 L_y}{2 \mu G_N}\sqrt{\frac{V_1}{pUB}} \right|_{z=1} \log\left(\frac1{1-z_*}\right) + \cdots ,\\
l &\approx \left.\mu \sqrt{\frac{V_2}{ p U B }} \right|_{z=1}
\log\left(\frac1{1-z_*}\right) + \cdots ,
\end{aligned}
\end{equation}
where $B \equiv 4 V_1 V_2 -V_1' V_2 - V_1 V_2'$ and
$p=1+z+z^2-\mu^2z^3/2 $.
Thus for large $l$,
\begin{equation} \label{asy}
S = r V_{\rm strip}+ \cdots, \quad {\rm with} \quad r
 = \left. \frac{\sqrt{V_1 V_2}}{2 G_N \mu^2} \right|_{z=1},
\end{equation}
is proportional to the volume $V_{\rm strip} = l L_y$ of the
strip, with $r$ the entropy density. Note that $S$ is given by the
horizon area and thus coincides with the black hole entropy. Thus
in the large distance limit, $S$ is solely determined by the
horizon geometry, which is of course expected. Note that the
result is in fact general, applying to generic gravity geometry
and entangling region of any shape (not just strip).

\begin{figure}
\center{
\includegraphics[scale=0.7]{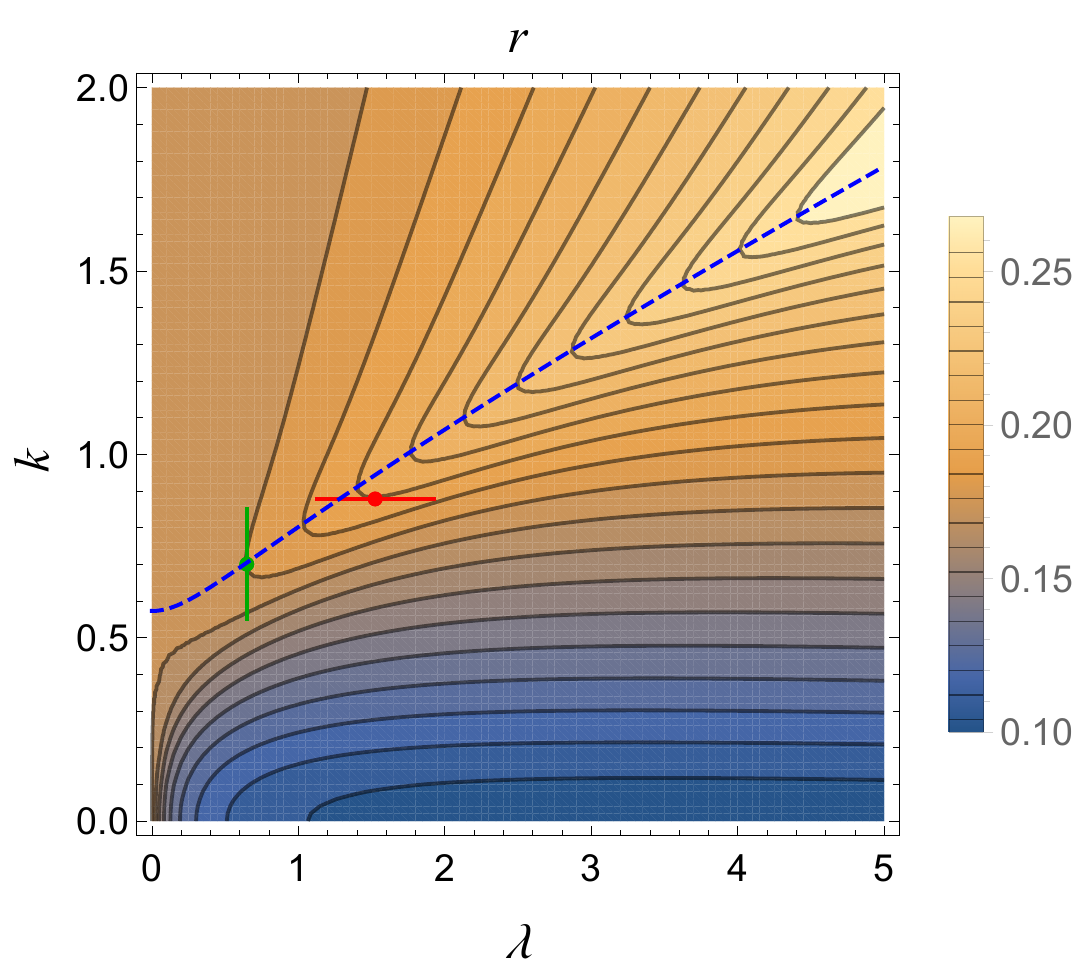}
\caption{\label{contour} A contour plot of the $r$, whose values
are represented by different colors. Each grey trajectory has the
same value of $r$. The blue dashed curve is the critical line
where metal-insulator transition occurs.}}
\end{figure}

In Fig.~\ref{contour} we give the contour plot of $r$ over the
$(\lambda,k)$-plane. Again, we see the $r$ achieves local maxima
precisely at the critical line identified from the DC conductivity
earlier. In particular, we show an example in this figure to
locate the maximal value of $r$ along the vertical direction (a
short green line ) and the horizontal direction (a short red
line), which are marked by a green dot and a red dot,
respectively. In general we find the data in this method matches
rather closely with Fig.~\ref{phase}. In other words, in long
distance limit entanglement entropy can also be used as a
diagnostic for the critical line of QPTs. Moreover, in this
contour plot we observe that the peaks of HEE form a ridge along
the direction of the critical line. Thus, when we change a single
parameter to locate the position of HEE peaks over two or higher
dimensional phase diagram, the result obviously depends on the
direction of observation (or the direction of cutting ridge).
Unlike in our current paper, in general cases the local HEE peaks
obtained in this manner may deviate far from critical points in
certain directions. Therefore, to avoid this confusion, we may
present a more strict statement on the relation between HEE and
QPTs for a multi-parameter system. That is, the peaks of HEE
always form a ridge along the critical line, and a pronounced peak
can be observed at the critical point when computing HEE along the
direction perpendicular to the critical line over the phase
diagram.

Our result for holographic systems resonates well with earlier
observations~\cite{Osborne:2002zz,Osterloh:2002na,Vidal:2002rm,
YChen:2006jop} in the condensed matter literature regarding
entanglement entropy and QCPs.

Furthermore, we have investigated the behavior of $r$ at lower
temperatures. We find that with the decrease of temperature, the
location of the maximum over the phase space tends to fix. In
metallic phase with fixed parameters the quantity $r$ converges to
non-zero values. While in insulating phase $r$ tends to be
vanishing in zero temperature limit, which implies the violation
of the volume law of HEE. As a result, the peaks of HEE become
steep on the insulating side with the decease of the temperature.
This phenomenon suggests that the maximization behavior of HEE at
finite temperature might be the reflection of the discontinuity of
HEE at zero temperature. However, as disclosed in
\cite{Donos:2012js}, in this case the quantum phase transition
could still be continuous and infinite order since the free energy
in the bulk does not exhibit any discontinuous behavior even at
zero temperature. Finally, it is worthwhile to point out that
for insulating phases there could exist a kink $(r'(T_c)=0)$ at
$T_c \simeq 10^{-5}$ as described in \cite{Donos:2013eha}.
However, for $T<T_c$ the thermodynamically preferred solution
still exhibits insulating behavior and $r$ has the singular
behavior $\lim_{T\to 0}r\to 0$. Therefore for even lower
temperature, the phase structure illustrated in Figure \ref{phase}
will be stable.

We conjecture that the discontinuity of HEE at zero temperature
may result from the artifact of IR geometry $AdS_2$.  As pointed
out in \cite{Donos:2012js,Iqbal:2011ae}, the IR geometry $AdS_2$
will lead to the violation of the third law of thermodynamics due
to the non-vanishing entropy density at zero temperature, and may
therefore be an artifact of large $N$ limit of holography.
Nevertheless, we intend to argue that the maximization behavior of
HEE disclosed in our paper may not suffer from this artifact. In
addition to the near horizon limit analysis shown above, Fig.
\ref{trend} shows that HEE with finite $l$ with $lT\ll 1$
also exhibits maximal behavior at critical points, which is not
dictated by the near horizon geometry since the minimal surface
can be distant from the horizon. This phenomenon can be understood
as that the HEE with finite $l$ actually captures the quantum
phase structure at the energy level specified by $z_*$, namely the
bottom location of the minimal surface. Furthermore, the peaks of
HEE converge to fixed values near the QCP when $l$ becomes large
(see Fig.\ref{phase}). Therefore, in spite of the subtlety of IR
geometry $AdS_2$, we believe the maximization behavior of HEE at
QPTs should be genuine and universal in the holographic approach.

\section{Discussion}
Throughout this paper we have worked with a
scalar mass $m^2 = -2$ because in this case we find the numerical
analysis exhibits convergent behavior very well. Nevertheless, it
is worth to point out that in this setup the $AdS_2$ $BF$ bound is
violated such that new phases could be found at lower temperatures
\cite{Donos:2013eha}. Taking this into account, we have also
performed the calculation for the mass $m^2=-3/2$ which does not
violate $AdS_2$ $BF$ bound. We qualitatively obtain the same
behavior of the HEE, but with much higher precision and more grid
points.

As a summary, in this paper we have computed the HEE of a strip
geometry in four dimensional Q-lattice background, which exhibits
metal-insulator transitions in the dual field theory. We have
found that the behavior of the HEE in long distance physics is
dominantly determined by the near horizon geometry of Q-lattice.
More importantly, through numerical calculation we have
demonstrated that the HEE attains a maximum at the corresponding
QCPs. This fact reveals that HEE can be used as a diagnostic for
QPTs indeed. Based on what we have observed in Q-lattice setup, we
have following remarks and conjectures which should be crucial for
next investigations on QPT in holographic approach. Firstly, we
conjecture that the connection between HEE and QPT which is
captured by the near horizon geometry would be a universal feature
for general holographic models. Secondly, we further propose a
simple but elegant criteria for the occurrence of QPT in
holographic approach, that would be the existence of a maximum for
the area element of black hole horizon when changing the
parameters of the system in zero temperature limit. Finally,
it is crucial to explore the scaling behavior of the HEE around
QCPs in the zero temperature limit, which should be done in a
background which contains IR geometry dual to metallic phases with
vanishing entropy density, rather than $AdS_2$ with finite entropy
density. Our investigation on these topics is under progress.

{\it Acknowledgement} - We express our special thanks to Hong Liu
for dramatically improving the manuscript. We are also grateful to Xiaomei
Kuang, Weijia Li, Rene Meyer, Philip Phillips, Yu Tian, Yidun Wan,
Kun Yang, Hongbao Zhang for stimulating discussion and helpful
correspondence. This work is supported by the Natural Science
Foundation of China under Grant Nos.11275208, 11305018 and
11178002. Y.L. also acknowledges the support from Jiangxi young
scientists (JingGang Star) program and 555 talent project of
Jiangxi Province. J. P. Wu is also supported by Program for
Liaoning Excellent Talents in University (No. LJQ2014123).

\end{document}